\documentstyle[12pt]{article}
\newcommand{\be}{\begin{equation}}
\newcommand{\ee}{\end{equation}}
\newcommand{\osi}{$^{16}$O}
\newcommand{\caf}{$^{40}$Ca}
\newcommand{\cafe}{$^{48}$Ca}
\newcommand{\bea}{\begin{eqnarray}}
\newcommand{\eea}{\end{eqnarray}}
\newcommand{\nn}{\nonumber}
\newcommand{\bfp}{\mbox{\bf p}}

\begin{document}
\title{NN Correlations and Relativistic Hartree-Fock\\
in Finite Nuclei}
\author{R.\ Fritz and H.\ M\"{u}ther
\\\\
Institut f\"{u}r Theoretische Physik,\\ Universit\"{a}t T\"{u}bingen,\\
D-72076 T\"{u}bingen, Germany}
\maketitle
\pagestyle{empty}

\date{\today}

\begin{abstract}
Two different approximation schemes for the self-consistent solution of
the relativistic Brueckner-Hartree-Fock equation for finite nuclei are
discussed using realistic One-Boson-Exchange potentials.
In a first scheme, the effects of correlations are deduced
from a study of nuclear matter and parameterized in terms of an
effective $\sigma$, $\omega$ and $\pi$ exchange. Employing this effective
interaction relativistic Hartree-Fock equations are solved for finite
nuclei \osi , \caf\ and \cafe . In the second approach the effect of
correlations are treated in the Brueckner-Hartree-Fock approximation
directly for the finite nuclei, but the modifications of the Dirac
spinors in the medium are derived from nuclear matter assuming a
local-density approximation. Both approaches yield rather similar
results for binding energies and radii in fair agreement with
experimental data. The importance of the density dependent correlation
effects is demonstrated and different ingredients to the spin-orbit
splitting in the shell-model of the nucleus are discussed.
\end{abstract}
\pagestyle{empty}

%\pacs{21.60.Jz, 21.10.Dr, 21.10.Ft, 21.65.+f}

\clearpage
\pagestyle{headings}

\section{Introduction}

The various attempts to derive the bulk properties of nuclear systems
from realistic nucleon-nucleon (NN) interactions are confronted with
two major obstacles. The first one is the necessity to consider the
effects of NN correlations which are due to the strong short-range and
tensor components in a realistic NN interaction. The second one is of a
relativistic nature: the strong scalar-meson ($\sigma$) exchange part
required in realistic meson-exchange potentials \cite{holi,rupr}, gives
rise to a significant modification of the Dirac structure of nucleons
in the nuclear medium \cite{serot}. Therefore relativistic features
should be included in the many-body theory of nuclear systems, in order
to account for this effect.

The importance of the NN correlations is made obvious by the fact that
no binding energy of nuclear systems is obtained if these correlations
are ignored: A Hartree-Fock (HF) calculation employing
e.g.~a realistic One-Boson-Exchange (OBE) potential \cite{rupr}, which
fits NN scattering data, yields unbound nuclei.
Various methods have been developed to include the effects of
two-nucleon correlations. One possibilitiy
is the so-called Brueckner-Hartree-Fock (BHF) approximation. In this
approach one considers a Slater-determinant, which should be an
appropriate model wave-function for the nuclear system to be investigated.
Solving the Bethe-Goldstone equation yields an effective interaction,
the G-matrix, which depends on the bare NN interaction and the model
wave-function considered. The self-consistency condition of BHF now
requires that the model-wave function, which is needed to set up the
Bethe-Goldstone equation, is made identical to the solution of the HF
equations using the G-matrix as a kind of effective interaction.

This self-consistency problem is simplified for nuclear matter since
the translational symmetry of this infinite system requires plane waves
for the single-particle wave functions to built up the
Slater-determinant. For many years, however, the BHF self-consistency problem
has also been solved for finite nuclei \cite{davies,tripat}.

The inclusion of NN correlations led to a substantial improvement in the
microscopic description of bulk properties of nuclei. For both nuclear
matter as well as finite nuclei BHF calculations employing various
realistic NN interactions gave results, which were located on the
so-called Coester band \cite{coester,mu1}. This means that either the
calculated binding energy turned out to be too small or the calculated
radii were too small (which corresponds to a saturation density of
nuclear matter too large) as compared to the experimental data. Attempts
have been made to improve the many-body approach such that results "off
the Coester band", closer to the experimental data were obtained. Such
attempts have been made within the hole-line expansion of the Brueckner
theory \cite{day} or using different schemes\cite{pand,zabo,rev}. Even
today it is not really clear if it is possible to derive the bulk properties
of nuclear systems from realistic NN forces within a non-relativistic
many-body theory \cite{kuo1}.

Motivated by the success of the phenomenological $\sigma$ - $\omega$
model of Walecka and Serot~\cite{serot}, attempts have been made to
incorporate the relativistic features of this approach also in nuclear
structure calculations which are based upon realistic NN forces.
In this approach, one accounts for the fact that the relativistic
nucleon self-energy in a nuclear medium is given essentially
by a large attractive component, originating mainly from the exchange
of the scalar $\sigma$ meson and therefore transforming like a scalar
under a Lorentz transformation, and a repulsive component, which
transforms like the zero component of a Lorentz vector and is mainly
due to the exchange of the $\omega$ meson. The single-particle motion
is described by a Dirac equation which includes this self-energy. The
strong components of the self-energy yield solutions of the Dirac
equation, which are quite different from the Dirac spinors describing
the nucleons in the vacuum.

This change of the Dirac spinors in the medium gives rise to a
self-consistency problem beyond the one already discussed above:
Already in evaluating the matrix elements of the bare NN force $V$ one
should know the structure of the Dirac spinors, resulting from the
solution of the Dirac equation. Again, this self-consistency problem is highly
simplified in nuclear matter. In this case the medium-dependence of the
Dirac spinors is characterized by an effective mass, which represents
the ratio of the small to the large component of the spinor.

Such Dirac BHF (DBHF) calculations
have been performed for nuclear matter by, e.~g., Shakin and
collaborators~\cite{shakin}, Brockmann and Machleidt~\cite{BM84},
and ter Haar and Malfliet~\cite{malf}. The basic aspects of this approach
have been thoroughly investigated by Horowitz and Serot~\cite{HS84}.
Due to the scalar field, the
nucleon mass is reduced enhancing the ratio between small
and large components of the Dirac spinors. This change in the Dirac
spinors yields a reduction of the scalar density, which implies that
the attraction due to the exchange of the $\sigma$ meson in OBE
potentials is reduced. At small densities of nuclear matter this loss
of attraction is counterbalanced by a reduction of the kinetic energy,
which is also caused by the medium dependence of the Dirac spinors. At
larger densities the loss of attraction in the NN interaction
overwhelms the loss of repulsion in the kinetic energy and for those
densities the energy calculated in the DBHF approximation is less
attractive than the corresponding energy calculated in the BHF
approximation ignoring these relativistic effects.

Consequently the saturation points calculated for nuclear matter in DBHF
approximation are shifted to smaller densities as compared to the BHF
result. Brockmann and Machleidt succeeded in constructing a realistic
OBE potential which fits NN scattering data and also yields DBHF
results for nuclear matter in satisfying agreement with the empirical
data \cite{BM84}. The same feature is also observed for the potential
$"A"$, defined in table A.2 of ref.\cite{rupr}, which we will consider
also in our present investigation.

This success of the DBHF approximation in nuclear matter gives rise to
the hope that the same DBHF approximation may also be successful to
reproduce the binding energies and radii of finite nuclei. From the
discussion above, it is obvious, however, that a complete
self-consistent calculation for finite nuclei is rather involved.
Therefore we are going to investigate two approximations, in which
either the effects of correlations or the relativistic effects are
taken from studies of nuclear matter, while the respective other
components of the calculation are treated in a self-consistent way
directly for the finite nuclei.

In the first approximation, we determine an effective meson theory
($\sigma$, $\omega$ and $\pi$ mesons), which yields in a Hartree-Fock
approximation for nuclear matter at a given density $\rho$ the same
observables for the self-energy of the nucleons and the binding energy
as a DBHF calculation of nuclear matter. This leads to a set of
coupling constants, which depend on the nuclear density. This density
dependence reflects the density dependence of the correlations
described by the G-matrix of DBHF. Keeping track of the density
dependence of these coupling constants one can perform a relativistic
HF calculation using techniques as described e.g.~by Bouyssy et.al.
\cite{bouys}. A calculation along this line has been performed by
Brockmann and Toki \cite{brto} restricted to a Hartree description and
first results of a HF calculation have been presented in \cite{fri1}.
Both of these earlier investigations allowed for an exchange of effective
$\sigma$ and $\omega$ mesons, only.

In the second approximation one is expanding the single-particle wave
functions of a self-consistent BHF calculation for finite nuclei in a basis
of plane waves. The Dirac structure of these plane waves is taken from
the corresponding state in nuclear matter of a density, which is equal to
the average density for the single-particle orbit in the finite nucleus
under consideration. In this way one deduces the Dirac effects from
nuclear matter, but otherwise performs a complete self-consistent BHF
calculation. This approach has also been used in ref.\cite{mmb1}.

It turns out that these two very different approximations yield rather
similar results. Therefore one can assume that a result of complete
DBHF calculation should also be close. It is the main aim of the
present investigation to explore the differences between the two
approaches in a systematic way. For that purpose we are studying
results on various nuclei (\osi , \caf , \cafe  ) using different OBE
interactions (OBE potentials "A" and "C" defined in table A.2 of
\cite{rupr}). After this introduction we define the details of the two
approximations towards a self-consistent DBHF calculation for finite
nuclei in sections 2 and 3, respectively. The results are presented and
discussed in section 4 and the main conclusions are summarized in section
5.

\section{Effective Meson Exchange Approach}

Our starting point for the description of the nuclear many -
body - problem is the effective lagrangian density for the
interacting nucleons and the $\sigma$, $\omega$ and $\pi$ meson
\be
    \cal L = \cal L_{\rm 0} + \cal L_{\rm I}  \; ,
\ee
consisting of the free lagrangian density
\bea
\cal L_{\rm 0} & = & \overline{\Psi} (i \gamma_{\mu} \partial^{\mu}
                   - M) \Psi
                   + \frac{1}{2} (\partial_{\mu} \Phi_{\sigma} \partial^{\mu}
                     \Phi_{\sigma} - m_{\sigma}^2 \Phi_{\sigma}^2 ) \nn \\
             &   & + \frac{1}{2} m_{\omega}^2 \Phi_{\omega,\mu}
\Phi_{\omega}^{\mu}
		     -\frac{1}{4} F_{\mu \nu} F^{\mu \nu}
                   + \frac{1}{2} (\partial_{\mu} \Phi_{\pi} \partial^{\mu}
                                  \Phi_{\pi} - m_{\pi}^2 \Phi_{\pi}^2 )
\eea
with
\be
    F_{\mu \nu} = \partial_{\mu} \Phi_{\omega ,\nu}
			    - \partial_{\nu} \Phi_{\omega ,\mu} \; ,
\ee
and the interaction lagrangian density
\bea
\cal L_{\rm I} & = & -G_{\sigma} \overline{\Psi} \Phi_{\sigma} \Psi
		   -G_\omega \overline{\Psi} \gamma_{\mu} \Phi_\omega^{\mu}
			\Psi
                   -\frac{f_{\pi}}{m_{\pi}} \overline{\Psi} \gamma_5
		   \gamma_{\mu} (\partial^{\mu} \Phi_{\pi})  \Psi \; .
\eea

The nucleon field and rest mass is denoted by $\Psi$ and M,
whereas the meson fields, rest masses and effective nucleon-meson
coupling constants are denoted by $\Phi_i$, $m_i$ and $G_i$ or $f_i$
with $ i=\{\sigma, \omega ,\pi\} $ for the scalar, vector and
pseudoscalar meson
respectively. Note that for the pion, we use the pseudovector coupling
and suppress the notations for the isospin degrees of freedom. Moreover
we already mention here, that we will subtract the zero range part in
the one-pion-exchange contributions to the NN interaction. This is done
to account for the effects of short-range correlations between the
interacting nucleons.

\subsection{Nuclear Matter}

Following standard techniques \cite{serot,bouys}, the Hartree-Fock
approximation for this lagrangian leads to a Dirac equation for
nucleons with four-momentum $p=(p^0,\bfp)$ in nuclear matter
\be
   [ \mbox{$\bf \gamma \cdot p$} + M + \Sigma (p) ] \Psi(\bfp,s) =
   \gamma_0 E(\bfp) \Psi(\bfp,s)  \; , \label{dirac}
\ee
for the nucleon spinors $ \Psi(\bfp,s)$, containing the self energy
$\Sigma (p)$. Because of the isotropy of nuclear matter the spinors
$ \Psi(\bfp,s)$ are known to be plane waves and in the rest frame of
nuclear matter the self energy $\Sigma (p)$ for on-shell
nucleons $\left(p^0=E(\bfp) \right)$ depends only
on the absolute value of the three-momentum $\bfp$.
This nucleon self energy can be split into different parts with a
well-defined behavior under Lorentz transformations. Because of
parity conservation, time reversal invariance and
hermiticity  the most general form of $\Sigma (\bfp)$ is restricted to
\be
    \Sigma (\bfp) = \Sigma^s (\bfp) - \gamma^0 \Sigma^0 (\bfp)
		   + \mbox{$\bf \gamma \cdot p$} \Sigma^v (\bfp)
\ee
with $\Sigma^s (p)$, $\Sigma^0 (p)$ and $\Sigma^v (p)$ transforming like
Lorentz scalars. Therefore the Dirac equation can be rewritten as
\be
   [ \mbox{$\bf \gamma \cdot p$}^* + M^*(\bfp) ] \Psi(\bfp,s) =
   \gamma_0 E(\bfp)^* \Psi(\bfp,s) \; ,
\ee
introducing the definitions
\bea
    \bfp^* &= \bfp \left(1+ \Sigma^v (\bfp) \right) \; ,\nn\\
    M^*(\bfp) &= M + \Sigma^s (\bfp) \; , \nn\\
    E^*(\bfp) &= E(\bfp) + \Sigma^0 (\bfp)  \; .\label{eq:staqu}
\eea
The formal similarity with a free Dirac equation allows immediately
to determine the nucleon spinor in nuclear matter to
\be
   \Psi(\bfp,s) = \left( \frac{E^*(\bfp) + M^*(\bfp)}{2 E^*(\bfp)}
\right)^{1/2}   \left( \begin{array}{c}
		       1 \\
		 \frac {\bf \sigma \cdot p^*} { E^*(\bfp) + M^*(\bfp) }
                      \end{array} \right)
               \chi^s \; ,       \label{spin}
\ee
now with a modified ratio of the spinors upper and lower component if
compared with the vacuum solution.
The spinors are normalized (noncovariant) to
\be
   \Psi(\bfp,s)^{\dagger} \Psi(\bfp,s) = 1 \; , \;
   \overline \Psi (\bfp,s) \Psi(\bfp,s) = \frac{M^*(\bfp)}{E^*(\bfp)}
\ee
and the on-shell condition in nuclear matter now reads like
\be
   E^*(\bfp)^2 = M^*(\bfp)^2 + {\bfp^*}^2 \; .
\ee
On the level of the Hartree-Fock approximation, the mesons used
in our lagrange density gives rise to the following contributions
to the self energy:
\bea
 \Sigma^s (\bfp) & = & \mbox{} -\left( \frac{G_\sigma}{m_\sigma} \right)^2
\rho_s \nn \\
		 &   & \mbox{} + \frac{1}{(4 \pi)^2} \frac{1}{p}
			 \int_0^{k_F} q dq \frac{M^*(q)}{E^*(q)}
                         \biggl[ G_\sigma^2 \Theta_\sigma (p,q)
			 - 4 G_\omega^2 \Theta_\omega (p,q) \nn \\
                 &   &  \mbox{}-3 \left( \frac{f_{\pi}}{m_{\pi}} \right)^2
			m_{\pi}^2 \Theta_{\pi} (p,q) \biggr] \\ \label{eq:nmhf1}
                 &   &  \nn \\
 \Sigma^0 (\bfp) & = & \mbox{} - \left( \frac{G_\omega}{m_\omega} \right)^2
\rho \nn \\
		 &   & \mbox{} - \frac{1}{(4 \pi)^2} \frac{1}{p}
			 \int_0^{k_F} q dq \biggl[ G_\sigma^2 \Theta_\sigma (p,q)
			 + 2 G_\omega^2 \Theta_\omega (p,q) \nn \\
                 &   &  \mbox{} - 3 \left( \frac{f_{\pi}}{m_{\pi}} \right)^2
			  m_{\pi}^2 \Theta_{\pi} (p,q) \biggr] \\ \label{eq:nmhf2}
                 &   &  \nn \\
 \Sigma^v (\bfp) & = & \mbox{} - \frac{1}{(4 \pi p)^2} \int_0^{k_F} q dq
                         \frac{q^*}{E^*(q)} \biggr[ 2 G_\sigma^2
\Gamma_\sigma (p,q)
                         + 4 G_\omega^2 \Gamma_\omega (p,q)  \nn \\
                 &   & \mbox{} - 6 \left( \frac{f_{\pi}}{m_{\pi}} \right)^2
                        \left( (p^2+q^2) \Gamma_{\pi} (p,q)
                        - pq \Theta_{\pi} (p,q) \right) \biggr]
          \label{eq:nmhf3}
\eea
We omit in our notation the obvious dependence of the self energies on
the Fermi momentum $k_F$. The first term in $\Sigma^s (\bfp)$ and
$\Sigma^0 (\bfp)$ corresponds to the Hartree contribution using
\be
     \rho (k_F) =  \frac{2}{3 \pi^2} k_F^3  \; \; \; {\rm and} \; \; \;
     \rho_s (k_F) =  \frac{2}{\pi^2} \int_0^{k_F} q^2 dq
			\frac{M^*(q)}{E^*(q)}
\ee
for the baryon and scalar density, respectively. The remaining expressions
are due to the Fock (exchange) contributions where we have used the
abbreviations
\bea
     A_i(p,q) & = & p^2 + q^2 + m_i^2 - \left( E(p)-E(q) \right)^2 \\
\Theta_i(p,q) & = & ln \left( \frac{A_i(p,q) + 2 pq}
				   {A_i(p,q) - 2 pq} \right) \\
  \Gamma_i(p,q) & = & \frac{A_i(p,q) \Theta_i(p,q)}{4pq} -1  \; ,
\eea
again $ i=\{\sigma ,\omega ,\pi\} $. Two important things have to be
noted. First, as already mentioned above, we have subtracted  zero-range
contributions from pion exchange to the
self energy $\Sigma (\bfp)$. Secondly, we don't want to consider retardation
effects in the meson propagators. Although retardation causes no
problems in nuclear
matter, the neglection leads to significant simplifications in finite nuclei.

In the next step we determine effective coupling constants $G_{\sigma}$
and $G_{\omega}$ for the scalar
and vector meson by requesting that the HF expressions for the  scalar
self energy $\Sigma^s (\bfp)$ calculated at the Fermi surface
($p=k_{f}$) and the binding energy per nucleon
reproduce the corresponding results of a
Dirac-Brueckner-HF (DBHF) calculation \cite{rupr,BM84} using realistic
NN-forces, namely versions A and C of the Bonn potential \cite{rupr}.
For the Pion  we fix the coupling constant to the free value
$f^2_{\pi}/ 4 \pi = 0.08 $ and the masses of the mesons are chosen to be
identical to those of the OBE potential ($m_{\pi}$=138 MeV,
$m_{\sigma}$=550 MeV and $m_{\omega}$=783 MeV). In this way we obtain
for each baryon density $\rho$ two effective coupling constants
$G_{\sigma}(\rho )$ and $G_{\omega}(\rho )$. The density dependence of
these coupling constants reflects the density dependence of the
correlations taken into account in the DBHF approximation.

\subsection{Finite Nuclei}

Once the density dependent coupling constants are determined, we assume
for the study of finite nuclei, that we can account for the density
dependent correlation effects in a relativistic HF calculation
by employing the coupling constants
calculated at the local density $G_{\sigma}(\rho (r))$ and
$G_{\omega}(\rho (r))$, where the density profile $\rho (r)$ is
determined from the result of the relativistic HF calculation in a
self-consistent manner. For that purpose we write the nucleon spinor for
the finite system in
coordinate space
\bea
   <r|\alpha> =  \Psi_{\alpha} (\vec{r}) & = &
      \left( \begin{array}{c}
	      g_a(r)  \\
	      -i f_a(r) \; \vec{\sigma} \cdot \hat{r}
	      \end{array}   \right)
      {\cal Y}_{\kappa_a, m_a} (\Omega) \;
      \chi_{\frac{1}{2},(q_a)} \\
      & = & \left( \begin{array}{c}
	    g_a(r) \; {\cal Y}_{\kappa_a, m_a} (\Omega) \\
	    i f_a(r) \; {\cal Y}_{-\kappa_a, m_a} (\Omega)
		   \end{array}   \right) \chi_{\frac{1}{2},(q_a)} \; .
\eea
Again, the spinors are normalized to
\be
   \int d^3 \vec{r} \Psi^{\dagger} ( \vec{r})
		    \Psi ( \vec{r})
   = \int_0^{\infty} r^2 dr \left[ g_a^2 (r) + f_a^2 (r) \right] = 1 \;
{}.
\ee
All quantum numbers are summarized by the index $\alpha=\{a,m_a\}$ with
$a=\{n_a,\kappa_a,q_a\}$. $n_a$ characterizes the radial quantum numbers,
whereas $\kappa_a=(2j_a + 1)(l_a - j_a)$ describes the angular momenta.
Obviously the upper and lower spinor-component for the same total angular
momentum $j_a$ have  different orbital
quantum numbers $l_a$. We introduce the corresponding $l_a'$ to the
same $j_a$
\be
   l_a' = \left\{ \begin{array}{l}
		  l_a + 1 \; {\rm for} \; l_a=j_a - \frac{1}{2} \\
		  l_a - 1 \; {\rm for} \; l_a=j_a + \frac{1}{2}
		  \end{array} \right.
\ee
and $a'=\{n_a,\kappa_a',q_a\}$, $\kappa_a'=(2j_a + 1)(l_a' - j_a)$.
${\cal Y}_{\kappa_a, m_a} (\Omega)$ is constructed as usual
\be
    {\cal Y}_{\kappa_a, m_a} (\Omega) =
    \sum_{m_l, m_s} (l_a m_l \; 1/2 \, m_s | j_a m_a)
    Y_{l_a, m_l} (\Omega) \chi_{\frac{1}{2}, m_s} \; .
\ee
For the isospinor $\chi_{\frac{1}{2},(q_a)}$ we use
\be
    q_a = \left\{ \begin{array}{l}
		  +1 \; {\rm for \: protons} \\
		  -1 \; {\rm for \: neutrons}
		  \end{array} \right.  \; .
\ee
The Dirac equation is solved by expanding the radial functions
$g_{a}(r)$ and $f_{a}(r)$ in a discrete basis of spherical Bessel
functions. The wave numbers for this basis are chosen such that this
discrete basis is a complete orthonormal basis in a sphere of radius
$D$. This radius is chosen to be large enough that the results for the
bound single-particle states are independent on $D$. With this
expansion the Dirac equation (\ref{dirac}) is rewritten in form of an
eigenvalue problem and the eigenvalues ($E_{a}$) and eigenvectors (the
expansion coefficients for $g_{a}$ and $f_{a}$) are determined by a
simple matrix diagonalisation \cite{thesis}.

In the following we give the expressions for the matrixelements of
the self-energy in this Dirac matrix, calculated in the Hartree-Fock
approximation. In this work we consider
nuclei with a closed proton and neutron shell only.
Therefore, the isovector pseudoscalar meson yields no contributions
in the Hartree approximation. The Hartree
matrix elements for the isoscalar scalar and vector part of the interaction
and the Coulomb force are given as
\bea
     <\alpha | \Sigma_\sigma^H | \beta> & = & -
     \delta_{\kappa_a,\kappa_b} \delta_{m_a,m_b} \delta_{q_a,q_b}
     \int_0^{\infty} r^2 dr G_{\sigma}(r)
     \left[ g_a (r) g_b (r) - f_a (r) f_b (r) \right]   \nn \\
 &  &\times\left\{ m_\sigma \int_0^{\infty} r'^2 dr' G_{\sigma}(r')\rho_s(r')
	    \widetilde{I}_0 (m_\sigma r_<) \widetilde{K}_0
         (m_\sigma r_>) \right\} \label{eq:hart1}\\
 &  &  \nn \\
     <\alpha | \Sigma_\omega^H | \beta> & = &
     \delta_{\kappa_a,\kappa_b} \delta_{m_a,m_b} \delta_{q_a,q_b}
     \int_0^{\infty} r^2 dr G_{\omega}(r)
     \left[ g_a (r) g_b (r) + f_a (r) f_b (r) \right]   \nn \\
 &  &\times \left\{  m_\omega \int_0^{\infty} r'^2 dr' G_{\omega}(r')\rho (r')
	    \widetilde{I}_0 (m_\omega r_<) \widetilde{K}_0 (m_\omega r_>) \right\}
\label{eq:hart2} \\
 &  &  \nn \\
     <\alpha | \Sigma_c^H | \beta> & = &
     \delta_{\kappa_a,\kappa_b} \delta_{m_a,m_b}
     \int_0^{\infty} r^2 dr
     \left[ g_a (r) g_b (r) + f_a (r) f_b (r) \right]   \nn \\
 &  &\times \left\{ e^2 \int_0^{\infty} r'^2 dr'
		 \frac{\rho_{p}(r')}{r_>} \right\}
     \frac{(1+q_a)}{2} \frac{(1+q_b)}{2} \label{eq:hart3}
\eea
with the definitions
\bea
    \rho (r)& = &\frac{1}{4 \pi} \sum_a \hat{\jmath}_a^2
		\left[ g_a^2 (r)  + f_a (r)^2  \right] \nn\\
    \rho_s(r)& = &\frac{1}{4 \pi} \sum_a \hat{\jmath}_a^2
		\left[ g_a^2 (r)  - f_a (r)^2  \right]
\eea
for the baryon and scalar density in finite nuclei, where
$\hat{\jmath} = \sqrt{2\jmath+1}$ and index a running over all
occupied orbits. If the baryon density of the protons $\rho_{p}(r)$
is needed as in the Coulomb self energy $\Sigma_c^H$, index a runs only over
occupied proton orbits. Note the use of the coupling constants
depending on the local position, which is just an abbreviation for
$G_{i}(r)=G_{i}(\rho (r))$. In eqs.(\ref{eq:hart1}) - (\ref{eq:hart3})
$r_<$ and $r_>$ are denoting the smaller or bigger value of $r$ and $r'$.
The functions $\widetilde{I}_L (x)$ and $\widetilde{K}_L (x)$ arise
from the multipole expansion of the meson propagator in coordinate space
and are defined using the modified spherical Bessel functions $I$ and
$K$\cite{abramo}:
\be
    \widetilde{I}_L (x) = \frac{I_{L+1/2} (x)}{\sqrt{x}} \; \; \: \:
    \widetilde{K}_L (x) = \frac{K_{L+1/2} (x)}{\sqrt{x}} \; .
\ee
The Fock contributions originating from the $\sigma$ exchange read
\bea
\lefteqn{ <\alpha | \Sigma_\sigma^F | \beta> =
          \delta_{\kappa_a,\kappa_b} \delta_{m_a,m_b} \delta_{q_a,q_b}
          \frac{(-m_\sigma)}{\hat{\jmath}_a^2} \sum_c \delta_{q_a,q_c}
           } \nn \\
&   &  \times   \int_0^{\infty} r^2 dr \biggl\{ G_{\sigma}(r)
          \left[ g_a (r) g_c(r) - f_a (r) f_c(r) \right]  \nn \\
&   & \times    \int_0^{\infty} r'^2 dr' G_{\sigma}(r')
          \left[ g_c(r') g_b(r') - f_c(r') f_b(r') \right]  \nn \\
&   & \times    \sum_L <a||Y_L||c>^2
          \widetilde{I}_L (m_\sigma r_<) \widetilde{K}_L
         (m_\sigma r_>) \biggr\}
\eea
with index c running over all occupied states. Here and in the following
expressions for the other mesons the $<a||Y_L||b>$ represent the
reduced matrix elements of the spherical harmonics and
\be
    <a||T_J(L)||b> = \sqrt{\frac{6}{4 \pi}} (-)^{l_a} \hat{l_a} \hat{l_b}
                     \hat{\jmath}_a \hat{\jmath}_b \widehat{L} \widehat{J}
                     \left( \begin{array}{c c c}
                            l_a & L & l_b \\
                             0  & 0 &  0
                            \end{array} \right)
                     \left\{ \begin{array}{c c c}
                             \jmath_a    & \jmath_b     & J \\
                               l_a       &   l_b        & L \\
                             \frac{1}{2} & \frac{1}{2}  & 1
                             \end{array} \right\}
\ee
Using this notation, the matrix elements of the Fock contributions
arising from the $\omega$ exchange can be written
\bea
<\alpha | \Sigma_{\omega}^F | \beta> & = &
          \delta_{\kappa_a,\kappa_b} \delta_{m_a,m_b} \delta_{q_a,q_b}
          \frac{m_\omega}{\hat{\jmath}_a^2} \sum_c \delta_{q_a,q_c}
           \nn \\
&   & \times \biggl\{   \int_0^{\infty} r^2 dr  G_{\omega}(r)
          \left[ g_a (r) g_c(r) + f_a (r) f_c(r) \right]  \nn \\
&   & \times    \int_0^{\infty} r'^2 dr' G_{\omega}(r')
          \left[ g_c(r') g_b(r') + f_c(r') f_b(r') \right]  \nn \\
&   &     \sum_L <a||Y_L||c>^2
        \widetilde{I}_L (m_\omega r_<) \widetilde{K}_L (m_\omega r_>) \biggr\}
\nn \\
& + & \sum_{L J} \biggl\{ \int_0^{\infty} r^2 dr G_{\omega}(r)
          \bigl[ g_a (r) f_c(r) <c'||T_J (L)||a>    \nn \\
&   &                     - f_a (r) g_c(r) <c||T_J (L)||a'> \bigr]
           \nn \\
&   & \times  \int_0^{\infty} r'^2 dr' G_{\omega}(r')
        \bigl[ g_c(r') f_b(r') <c||T_J (L)||a'> \nn \\
&   &        - f_c(r') g_b(r') <c'||T_J (L)||a> \bigr]
        \widetilde{I}_L (m_\omega r_<) \widetilde{K}_L (m_\omega
      r_>)  \biggr\}
\eea
The isovector pseudoscalar meson using the pseudovector coupling yields
\bea
\lefteqn{ <\alpha | \Sigma_{\pi}^F | \beta> =
          \delta_{\kappa_a,\kappa_b} \delta_{m_a,m_b} \delta_{q_a,q_b}
          \left( \frac{f_{\pi}}{m_{\pi}} \right) ^2
          \frac{1}{\hat{\jmath}_a^2} \sum_c (2 - \delta_{q_a,q_c})
 } \nn \\
&   & \times    \int_0^{\infty} r^2 dr  \biggl\{
          - \frac{\tilde{\jmath}_a^2 \tilde{\jmath}_b^2}{8 \pi}
          \left[ g_a (r) g_c(r) + f_a (r) f_c(r) \right]
          \left[ g_b (r) g_c(r) + f_b (r) f_c(r) \right]  \nn \\
&   &     + m_{\pi}^3 \sum_L \widehat{L} ^{-4}
          |<a||Y_L||c'>| ^2 \sum_{L_1 L_2} i^{L_2 - L_1}  \nn \\
&   &  \times   \left[ \left(\kappa_a + \kappa_c + h(L_1)\right) g_a (r) g_c(r)
             -\left(\kappa_a + \kappa_c - h(L_1)\right) f_a (r) f_c(r) \right]
           \nn \\
&   & \times    \int_0^{\infty} r'^2 dr'
          \left[ (\kappa_b + \kappa_c + h(L_2)) g_b (r) g_c(r)
                -(\kappa_b + \kappa_c - h(L_2)) f_b (r) f_c(r) \right]
           \nn \\
&   &    \times \hspace{2 cm} R (L_1,L_2,r,r') \nn \\
&   &     + \frac{1}{3} \sum_{L J}
	  \left[ <a||T_J(L)||c> g_a (r) g_c(r)
	       + <a'||T_J(L)||c'> f_a (r) f_c(r) \right] \nn \\
&   & \times    \hspace{1.2 cm}\left[ <a||T_J(L)||c> g_b (r) g_c(r)
             + <a'||T_J(L)||c'> f_b (r) f_c(r) \right] \biggr\}
\eea
with $L_i = \{L-1,L+1\}$ and the auxiliary functions
\be
    h(L_i) = \left\{ \begin{array}{c l}
                        -L & {\rm if} \; \; L_i=L-1 \\
                       L+1 & {\rm if} \; \; L_i=L+1
                     \end{array} \right. \; \; {\rm and}
\ee
\be
    R (L_1,L_2,r,r') =
    \theta(r'-r) \widetilde{I}_{L_1} (m_{ps} r)
                 \widetilde{K}_{L_2} (m_{ps} r')
  +\theta(r-r') \widetilde{I}_{L_2} (m_{ps} r')
                \widetilde{K}_{L_1} (m_{ps} r)
\ee
using the step function $\theta(x-y)$.
The expressions for the self energy show that the Hartree contributions
can be rewritten by defining a local potential, e.g.
\bea
     <\alpha | \Sigma_\sigma^H | \beta> & = &
     \delta_{\kappa_a,\kappa_b} \delta_{m_a,m_b} \delta_{q_a,q_b}
     \int_0^{\infty} r^2 dr
     \left[ g_a (r) g_b (r) - f_a (r) f_b (r) \right] V_\sigma (r)  \nn \\
& = & \delta_{\kappa_a,\kappa_b} \delta_{m_a,m_b} \delta_{q_a,q_b}
\int d^3r \Psi_{\alpha}^{\dagger} (r) \gamma^0 V_\sigma (r)
\Psi_{\beta} (r)
\eea
with
\be
    V_\sigma (r) = -G_{\sigma}(r) m_\sigma \int_0^{\infty} r'^2 dr'
\rho_s(r') G_{\sigma}(r')
            \widetilde{I}_0 (m_\sigma r_<) \widetilde{K}_0
               (m_\sigma r_>) \; \; ,
\ee
whereas the Fock contributions are obviously non-local.

Solving the Dirac equation (\ref{dirac}) with the technique mentioned
above in a self-consistent way one can finally determine the binding
energy of the nucleus with $A$ nucleons as
\be
E = \frac{1}{2} \sum_{\alpha, \mbox{( occ)}} E_{\alpha} + T_{\alpha}
\quad - AM \; ,
\ee
where $E_{\alpha}$ denotes the single-particle energy obtained by
solving the Dirac equation and $T_{\alpha}$ is the corresponding kinetic
energy.

\section{Local Density Approximation}

In contrast to the effective meson exchange approach, which, as discussed
in the preceding section, determines the effects of correlations from
studies of nuclear matter, we are now considering an approximation, in
which the relativistic effects are deduced from nuclear matter. For
that purpose we consider as an example the expansion of an harmonic
oscillator (h.o.) state in terms of plane wave spinors
\bea
   \lefteqn{ \Psi_{n,j,l,m} (\bfp,\rho ) =} \nn \\
&  & \sqrt{\frac{E^*(\bfp,\rho ) + M^*(\bfp,\rho )}{2 E^*(\bfp,\rho )}}
      \left( \begin{array}{c l}
                       1 & {\cal Y}_{j l m} (\Omega_p) \\
                \frac{  \sigma \cdot p^*(\rho )}
                     {  [ E^*(p,\rho ) + M^*(p,\rho ) ]}&
                       {\cal Y}_{j l' m} (\Omega_p)
                      \end{array} \right)
      \Phi_{n,l} (p) \label{ho}
\eea
where $\Phi_{n,l} (p)$ are the momentum space h.o. wavefunctions. The
structure of the plane wave Dirac spinor, in particular the ratio of the
small to large component is determined by the quantities $p^*$, $M^*$
and $E^*$ as defined in eq.(\ref{eq:staqu}). The values actually used
for these quantities are taken from the DBHF calculations at a density
$\rho$ for the realistic NN interaction under consideration. In this
sense the Dirac structure of the harmonic oscillator state defined in
eq.(\ref{ho}) is derived from nuclear matter of a given density $\rho$.
For the Dirac spinors as presented in eq.(\ref{ho}) one can calculate
the matrix elements of the OBE potential under consideration employing
the conventional techniques and identify the resulting numbers with
matrix elements of a two-body interaction $V$ between non-relativistic
harmonic oscillator states
\be
<\alpha \beta \vert V(\rho ) \vert \gamma \delta > \label{eq:vdef}
\ee
where $\alpha \dots \delta$ refer to the quantum numbers of the various
h.o. states and the parameter $\rho$ is kept to memorize that the value
of this matrix element depends on a density parameter $\rho$, which
determines the Dirac structure of the spinors used to calculate the
matrix element. This scheme can of course be generalized to
single-particle wave functions different from h.o. functions. In a
corresponding way one can also evaluate the matrix elements for the
operator of the kinetic energy
\bea
t_{\alpha\beta}(\rho ) & = \int d^3p \Psi_{\alpha}^\dagger \left[
\mbox{$\bf \gamma \cdot p$} + M\right] \Psi_{\beta} - M
\delta_{\alpha\beta} \nn \\
& = \int d^3p \Phi^*_{\alpha}\left[ \frac{M M^*(\rho ) + p
p^*(\rho )}{E^*(\rho)} - M \right] \Phi_{\beta} \label{eq:tkin}
\eea
For the interaction defined by the matrix elements of
eq.(\ref{eq:vdef}) one may now solve the Bethe-Goldstone equation
\be
G(Z,\rho ) = V(\rho ) + V(\rho ) \frac{Q}{Z-QH_{0}Q} G(Z, \rho )
\label{eq:betheg}
\ee
using the standard techniques of non-relativistic BHF calculations for
finite nuclei \cite{sauer}. The Pauli operator $Q$ in this equation is
defined in terms of h.o. states appropriate for the nucleus under
consideration. Beside the usual dependence on the starting energy $Z$,
the matrix elements of $G$ also depend on the density parameter $\rho$
characterizing the structure of the Dirac spinors involved. Keeping
track of this additional density dependence one can expand the BHF
single-particle states $\vert i >$ and $\vert j >$ in the basis of h.o.
states $\vert \alpha >$
\be
\vert i > = \sum_{\alpha} c_{i\alpha} \vert \alpha >
\ee
and the expansion coefficients are determined from the solution of the
eigenvalue problem
\be
\sum_{\beta} \left[ t_{\alpha\beta}(\rho_{i}) + \sum_{j,\mbox{( occ)}}
<\alpha j \vert G(Z=\epsilon_{i}+\epsilon_{j}, \rho_{ij})\vert \beta j>
\right] c_{i\beta}    = \epsilon_{i} c_{i\alpha} \label{eq:bhf}
\ee
If one ignores in this equation the medium dependence of the Dirac
spinors by putting $\rho_{i}$ and $\rho_{ij}$ equal to zero this
eq.(\ref{eq:bhf}) together with the Bethe-Goldstone
eq.(\ref{eq:betheg}) defines the conventional BHF approach for
realistic OBE potentials \cite{carlo}. In addition to the
self-consistency requirements of this BHF approach, we now want to
account for the medium dependence of the Dirac spinors and define an
average density for nucleons in the orbit $i$ by
\be
\rho_{i} = \int d^3r \; \phi_{i}^*(r) \rho_{DBHF}(r) \phi_{i}(r)
\label{eq:deni}
\ee
with $\phi_{i}(r)$ the DBHF single-particle wave function and
$\rho_{DBHF}(r)$ the radial shape of the baryon density obtained
from this calculation. This average single-particle density enters into
the calculation of the kinetic energy and it is also used to define the
average density for an interacting pair of nucleons by
\be
    \rho_{ij} = \sqrt{\rho_i \rho_j} \; \; .
\ee

\section {Results and Discussion}

As a first step towards the application of the effective meson exchange
approach discussed in section 2, we have to determine the coupling
constants $G_{\sigma}(\rho )$ and $G_{\omega}(\rho )$, depending on the
baryon density $\rho$. As discussed in subsection 2.1 this is done by
adjusting these parameters in such a way that a mean-field calculation
reproduces at each density the results for the scalar self-energy
$\Sigma^s$ and the binding energy per nucleon obtained in DBHF
calculations for nuclear matter \cite{brom}. For the mean-field
calculation we consider three different approximations. In the first
approach, we just consider the Hartree-contributions to the self-energy
and total energy (see eqs.(\ref{eq:nmhf1}) and (\ref{eq:nmhf2})). In
the second approach we consider a $\sigma$-$\omega$ model in
Hartree-Fock approximation, i.e.~we keep the Hartree and the Fock terms
in eqs.(\ref{eq:nmhf1}) to (\ref{eq:nmhf3}) which originate from the
exchange of a $\sigma$ or $\omega$ meson. This approach will be called
HF($\sigma,\omega$) or HF1. In the approach HF2 or
HF($\sigma,\omega,\pi$) we furthermore consider the effects of the
pion-exchange, which means that we are considering the complete set
outlined in section 2.

Results for the effective coupling constants determined from DBHF
calculations for OBE potentials A and C \cite{rupr,brom}, are listed in
table 1 for various densities. For all three approaches considered, the
effective coupling constants $G_{\sigma}$ and $G_{\omega}$ decrease
with increasing density. This is also displayed in fig.~\ref{fig:coupac}. The
decrease reflects the fact that the terms in the G-matrix, which are of
second and higher order in the interaction, contain  contributions,
which are simulated by the exchange of a scalar and a vector meson
\cite{elsen}. Due to the Pauli operator in the Bethe-Goldstone equation
(\ref{eq:betheg}) and due to the change in the energy dominator, these
contributions of higher order in the bare interaction $V$ are quenched
with increasing density. The importance of the density-dependent
correlation effects parameterized in terms of these coupling constants
is reflected by the fact that the square of the coupling constants are
quenched by a factor 2, if the nuclear density is increased from 0.2
$\rho_{0}$ to 1.4 $\rho_{0}$ ($\rho_{0}$ denoting the empirical
saturation density).

Such a substantial density dependence of the
effective coupling constants of course effects the nuclear structure
calculations. This is displayed in fig.~\ref{fig:nmhaa}, where the
results for the binding energy per nucleon and the effective mass of
the nucleon determined in a DBHF calculation are compared to the
corresponding quantities obtained in Hartree calculations using effective
coupling constants which are determined to reproduce the DBHF results
at a small density (dashed curve) or large density (dotted line). It is
obvious that the large coupling constant $G_{\sigma}$ determined for
small densities predicts an effective mass at larger density, which is
considerably smaller than the one obtained in the self-consistent DBHF
calculation. This is an example to demonstrate that mean field
calculations employing constant coupling constants tend to overestimate
the change of the Dirac spinors in nuclear matter at high densities.

The density dependent coupling constants also effect the predictions
for the saturation properties of nuclear matter. Note that the energy
versus density curve obtained from the Hartree calculation with
coupling constant appropriate for small densities yields a minimum at a
low density and overestimates the binding energy by about 8 MeV per
nucleon. The Hartree calculations using coupling constants for large
densities (dotted line in fig.~\ref{fig:nmhaa}) predict a minimum at higher
densities and a lower energy per nucleon. Therefore the minima of the
energy versus density curves obtained from coupling constants
determined at various densities form a ``band'' of saturation points,
which is perpendicular to the ``Coester band''.

The same phenomenon can also be observed in calculations of finite
nuclei. This is demonstrated in table~\ref{tab:hf2den} where we show
results of relativistic Hartree-Fock calculations considering coupling
constants as derived from various densities in nuclear matter.
The binding energies displayed in this table and all subsequent ones
are corrected for spurious center of mass effects,
assuming an harmonic oscillator model, and the radius of the charge
distribution has been evaluated from the proton density, assuming a radius
of 0.8 fm for the charge radius of the proton.

Inspecting the results displayed in table~\ref{tab:hf2den} one
observes a sensitive dependence of the calculated binding energy on the
choice for the coupling constants. This demonstrates again the
importance of the density dependent correlation effects contained in
the G-matrix of nuclear matter. It is striking to see that the
calculation which yields the largest binding energy
($\rho=0.2\rho_{0}$) also predicts the largest radius of the charge
distribution. So we find also for finite nuclei that the ground-state
properties of finite nuclei calculated with meson exchange parameters
derived from various densities form a ``band'' which is perpendicular
to the normal ``Coester band'' \cite{carlo}.

Furthermore its worth noting that the spin-orbit splitting deduced from
the difference in the single-particle energies $\epsilon_{p3/2}$ and
$\epsilon_{p1/2}$ is of course largest for that interaction which
yields the smallest effective mass in nuclear matter
($\rho=0.2\rho_{0}$). The comparison displayed in table~\ref{tab:hf2den}
demonstrates the importance of relativistic effects for the spin-orbit
splitting in the nuclear shell-model \cite{larry}.

Table~\ref{tab:hf2den} also contains a first result which is obtained
when we consider an effective meson exchange with local coupling
constants, depending on the position of the interacting nucleons as
discussed in section 2.2. As to be expected, one finds that the
single-particle energy for the $p_{1/2}$ state obtained in this
self-consistent calculation is closer to the one obtained
$\rho=0.2\rho_{0}$, while the single-particle energy of the deep lying
$s_{1/2}$ state is closer to the one obtained at larger densities. It
should be mentioned that effective coupling constants can safely be
derived from nuclear matter calculations only for densities as low as
$\approx 0.2 $ times the saturation density $\rho_{0}$. For smaller
densities the conventional tools to evaluate BHF energies yield
unstable results \cite{rev}. Therefore we extrapolate the coupling
constants to smaller densities using spline functions in terms of the
density.

The next question we want to investigate is the sensitivity of the
effective meson exchange model on the mesons taken into account.
For that purpose we have performed Hartree calculations for finite
nuclei using the local coupling constants as derived from the Hartree
calculations of nuclear matter (column ``Hart'' in table
\ref{tab:coup}). In the same way we also perform HF calculations for
finite nuclei, taking into account the effects of $\sigma$ and $\omega$
exchange using the density-dependent coupling constants derived from
nuclear matter (``HF1'') and finally consider the complete model with
inclusion of the pion discussed in section 2 (``HF2'').

Results obtained for these three effective meson exchange models for
the nuclei \osi , \caf , and \cafe\ are displayed in tables
\ref{tab:resosi}, \ref{tab:rescaf} and \ref{tab:rescafe}, respectively.
Using the OBE potential A, which yields a correct description of the
saturation point of nuclear matter, the Hartree approximation shows
fair agreement with the experimental data for the binding energy and
radius of all three nuclei considered. Both the results for the radius
and the binding energy are slightly below the experimental values.
Employing potential C yields larger radii but smaller binding energies.
This is the typical feature for 2 phase-shift equivalent potentials,
the results change along the ``Coester band''.

The inclusion of the Fock terms in HF1 reduces the calculated radii to
a significant extent, predicting the same or a slightly smaller binding
energy as compared to the Hartree approach. Therefore the agreement
with experiment gets worse. Furthermore we note that the Fock terms
tend to enhance the spin-orbit splitting in the single-particle
energies, which again deteriorates the agreement with the splitting
observed in the experimental data.

The pion-exchange terms included in the HF2 approximation
slightly improve the agreement between calculation and experiment. The
spin-orbit splitting is reduced and the binding energies are larger but
the results for the radii are essentially the same as in the HF1
approximation. Keeping in mind the sensitivity of the calculated
values on the density dependence of the effective coupling constants
displayed in table \ref{tab:hf2den} and discussed above, one may
conclude, however, that all three models lead to results, which are
rather similar.

The main purpose of this study is to compare the predictions of the
effective meson exchange approach to the results obtained in BHF
calculations, which treat the change of the Dirac spinors in the medium
in a local density approximation (see section 3 and ref.\cite{mmb1}).
For that purpose the tables \ref{tab:resosi} and \ref{tab:rescaf}
show results of this approach (identified as DBHF) and allow a
comparison with conventional BHF calculations, which ignore medium
dependence of the Dirac spinors completely. The differences between BHF
and DBHF results are by far not as large as those displayed in table
\ref{tab:hf2den}, which reflect  the density dependence of the
correlations. Therefore we conclude that the bulk properties of nuclei
are more sensitive to the density dependence of the correlations than
to the medium dependence of the Dirac spinors. That is why we consider
the approach, which treats the Dirac effects in a local density
approximation (DBHF), to give more reliable predictions for a complete
Dirac - Brueckner - Hartree - Fock calculation than the effective meson
exchange approach, which derives the correlation effects from nuclear
matter.

For the case of the nucleus \osi\ it has already been shown in
\cite{mmb1} that the inclusion of Dirac effects in DBHF leads to larger
radii and binding energies as compared to the predictions of
conventional BHF calculations. Thereby the agreement of the theoretical
predictions are substantially improved. This observation is supported
by the results on \caf\ shown in table \ref{tab:rescaf}.
Furthermore one finds that the DBHF results are in fair agreement with
those obtained in the relativistic HF approximation using effective
meson exchange. This agreement supports the conclusion that both types
of approaches are reliable approximation for a complete Dirac Brueckner
calculation.

Finally, we want to investigate the basic assumption of the DBHF
approach which assumes that the Dirac spinors for the single-particle
states in finite nuclei can be described in terms of plane wave spinors
of nuclear matter. For that purpose we consider as an example the
radial functions $g_{a}(r)$ and $f_{a}(r)$ for the large and small
component of the $0s_{1/2}$ Dirac spinor calculated in a relativistic
HF approach (HF1, OBEPA) for \caf\ (see solid lines in
fig.~\ref{fig:wave}). For the comparison we consider a Dirac spinor
expanded in terms of spinors for nuclear matter assuming a harmonic
oscillator expansion as in eq.(\ref{ho}). If we consider plane-wave
Dirac spinors of the vacuum ($k_{F}$=0, dotted line) the lower or small
component is considerably weaker than the one resulting from the
relativistic HF calculation. For the appropriate average density
$\rho_{i }$ as defined in eq.(\ref{eq:deni}) the enhancement of the
small component in the medium is fairly well described. This
demonstrates that the 2 approaches not only lead to very similar
results for the global observables like binding energy and radius, but
also provide similar predictions for the components of the Dirac
spinors.

\section{Conclusions}

Two different steps towards a self-consistent Dirac Brueckner
calculation for finite nuclei are presented and discussed. In the
effective meson exchange approximation one solves the relativistic
Hartree-Fock equations directly for the finite system and deduces the
correlation effects from nuclear matter. This is done in various models
to study the importance of Fock exchange effects and the impact of the
pion exchange. The density dependence of the effective coupling
constants reflects the density dependence of the correlations
encountered in the Brueckner G matrix.

In an alternative approach (DBHF) the correlation effects are treated
directly for the finite nuclei but the change of the Dirac spinors is
determined from nuclear matter. It is demonstrated that the bulk
properties of nuclei (binding energy and radius) are less affected by
the change of the Dirac spinors than by the density dependence of the
correlations. This implies that the approach which treats the
correlation effects without approximation should provide more reliable
results than the effective meson exchange approach.

It turns out that both approximations yield very similar results. For
the realistic OBE potential A defined in \cite{rupr} binding energies
per nucleon are obtained for \osi , \caf\ and \cafe , which are close
to the experimental value ($\pm$ 0.5 MeV). The predictions for the
radii are still significantly below the experimental data (typically
0.2 fm). This might be improved by including correlations beyond the
lowest order Brueckner theory. Recently it has been demonstrated that
the inclusion of hole-hole scattering terms within a self-consistent
Green function approach tends to improve BHF results in this direction
\cite{sko93}.

This work has partly been supported by the Graduiertenkolleg "Struktur und
Wechselwirkung von Hadronen und Kernen", T\"ubingen
(DFG, Mu705/3-1).

\clearpage
\pagestyle{plain}
%\section{Tables}

\begin{table}
\caption{DBHF results for nuclear matter derived from OBE potential $A$
for the scalar part of nucleon self-energy ($\Sigma^s$) and the binding
energy per nucleon ($E/A$) for various Fermi momenta $k_{f}$. The
column $G_{\sigma}$ and $G_{\omega}$ show the coupling constants which
are needed to reproduce these results in a HF calculation. For this
purpose three different models are considered: the Hartree
approximation ignoring the all Fock contributions to the self-energy
and binding energy (model ``Hart''), the Hartree-Fock approximation
ignoring the contribution of the pion-exchange (model ``HF1'') and the
full model defined in section 2.1 (model ``HF2''). All energies are
listed in MeV and the Fermi momenta in unit fm$^{-1}$. For a comparison
the lowest part of the table also shows results obtained for OBE
potential C at one specific density.}
\label{tab:coup}
\begin{center}
\begin{tabular}{||r|rr|c|rr||}
\hline\hline
&&&&&\\
\multicolumn{1}{||c}{$k_{F}$}&\multicolumn{1}{|c}{E/A}&\multicolumn{1}{c}
{$\Sigma^s$}&\multicolumn{1}{|c}{Model}&\multicolumn{1}
{|c}{$G_{\sigma}$} & \multicolumn{1}{c||}{$G_{\omega}$}\\
&&&&&\\
\hline
&&&&&\\
0.80 & -7.27 & -134.3 & Hart & 12.436 & 15.403 \\
&&& HF1 & 11.411 & 12.941 \\
&&& HF2 & 11.227 & 13.179 \\
&&&&&\\
1.00 & -10.62 & -209.8 & Hart & 11.177 & 13.807 \\
&&& HF1 & 10.265 & 11.668 \\
&&& HF2 & 10.104 & 11.885 \\
&&&&&\\
1.20 & -13.44 & -288.8 & Hart & 10.059 & 12.322 \\
&&& HF1 & 9.267 & 10.470 \\
&&& HF2 & 9.118 & 10.674 \\
&&&&&\\
1.40 & -15.59 & -374.9 & Hart & 9.224 & 11.168 \\
&&& HF1 & 8.531 & 9.539 \\
&&& HF2 & 8.389 & 9.733 \\
&&&&&\\
1.50 & -14.88 & -416.3 & Hart & 8.851 & 10.673 \\
&&& HF1 & 8.197 & 9.145 \\
&&& HF2 & 8.048 & 9.336 \\
&&&&&\\
\hline
\multicolumn{6}{||c||} { }\\
\multicolumn{6}{||c||} {OBE Potential C}\\
\multicolumn{6}{||c||} { }\\
1.20 & -11.57 & -292.8 & Hart & 10.130 & 12.534 \\
&&& HF1 & 9.297 & 10.669 \\
&&& HF2 & 9.149 & 10.869 \\
&&&&&\\
\hline\hline
\end{tabular}
\end{center}
\end{table}

\clearpage

\begin{table}
\caption{Results of relativistic HF calculations on
\osi , considering
the exchange of effective $\sigma$, $\omega$ and $\pi$ mesons. The
coupling constants are determined to reproduce DBHF results for nuclear
matter (OBE potential A) at various densities: $\rho=0.2\rho_{0}$
($k_{F}$=0.8 fm$^{-1}$), $\rho=0.5\rho_{0}$ ($k_{F}$=1.1 fm$^{-1}$) and
$\rho=1.4\rho_{0}$ ($k_{F}$=1.5 fm$^{-1}$). These results can be compared
to those of a self-consistent calculation (last column) considering local
coupling constants as discussed in section 2.2. Results are presented
for the single-particle energies of proton states, the binding energy
per nucleon (E/A, corrected for cm effects) and the radius of the charge
distribution ($R_{ch}$).}
\label{tab:hf2den}
\begin{center}
\begin{tabular}{||c|rrr|r||}
\hline\hline
&&&&\\
& \multicolumn{1}{|c}{$\rho = 0.2 \rho_{0}$}
&\multicolumn{1}{c}{$\rho = 0.5 \rho_{0}$}
&\multicolumn{1}{c}{$\rho = 1.4 \rho_{0}$}
&\multicolumn{1}{|c||}{self-cons.}\\
&&&&\\
\hline
&&&&\\
$\epsilon_{s1/2}$ [MeV] & -54.4 & -48.8 & -46.6 & -47.1 \\
$\epsilon_{p3/2}$ [MeV] & -33.9 & -25.4 & -20.1 & -23.8 \\
$\epsilon_{p1/2}$ [MeV] & -18.6 & -13.7 & -11.3 & -17.7 \\
&&&&\\
E/A [MeV] & -15.17 & -9.29 & -5.38 & -7.73 \\
$R_{ch}$ [fm] & 2.52 & 2.46 & 2.37 & 2.48 \\
&&&&\\
\hline\hline
\end{tabular}
\end{center}
\end{table}
\clearpage

\begin{table}
\caption{Results of relativistic HF calculations on
\osi , considering various models for the effective meson exchange
(Hart, HF1, HF2, see table 1) are compared to results of conventional BHF
calculations and BHF calculations which account for Dirac effects in
the way described in section 3 (DBHF). Further information see table
2.}
\label{tab:resosi}
\begin{center}
\begin{tabular}{||c|rrr|rr|r||}
\hline\hline
&&&&&&\\
& \multicolumn{1}{|c}{Hart.}
&\multicolumn{1}{c}{HF1}
&\multicolumn{1}{c}{HF2}
&\multicolumn{1}{|c}{BHF}
&\multicolumn{1}{c}{DBHF}
&\multicolumn{1}{|c||}{Exp}\\
&&&&&&\\
\hline
&\multicolumn{6}{|c||}{}\\
&\multicolumn{6}{|c||}{Potential A}\\
&\multicolumn{6}{|c||}{}\\
&&&&&&\\
$\epsilon_{s1/2}$ [MeV] & -44.0 & -44.0 & -47.1 & -56.6 & -49.8 &
-40$\pm$8 \\
$\epsilon_{p3/2}$ [MeV] & -21.5 & -23.4 & -23.8 & -25.7 & -23.0 & -18.4 \\
$\epsilon_{p1/2}$ [MeV] & -15.8 & -15.8 & -17.7 & -17.4 & -13.1 & -12.1 \\
&&&&&&\\
E/A [MeV] & -7.20 & -7.23 & -7.73 & -5.95 & -7.56 & -7.98 \\
$R_{ch}$ [fm] & 2.57 & 2.48 & 2.48 & 2.31 & 2.46 & 2.70 \\
&&&&&&\\
\hline
&\multicolumn{6}{|c||}{}\\
&\multicolumn{6}{|c||}{Potential C}\\
&\multicolumn{6}{|c||}{}\\
&&&&&&\\
$\epsilon_{s1/2}$ [MeV] & -37.0 & -37.4 & -40.2 & -45.2 & -40.9 &
-40$\pm$8 \\
$\epsilon_{p3/2}$ [MeV] & -17.7 & -19.4 & -19.7 & -19.5 & -18.0 & -18.4 \\
$\epsilon_{p1/2}$ [MeV] & -13.3 & -13.5 & -14.9 & -13.7 & -11.0 & -12.1 \\
&&&&&&\\
E/A [MeV] & -5.60 & -5.59 & -6.09 & -4.03 & -5.30 & -7.98 \\
$R_{ch}$ [fm] & 2.73 & 2.62 & 2.62 & 2.48 & 2.59 & 2.70 \\
&&&&&&\\
\hline\hline
\end{tabular}
\end{center}
\end{table}
\clearpage

\begin{table}
\caption{Results of relativistic HF calculations on
\caf . Further information see table 3.}
\label{tab:rescaf}
\begin{center}
\begin{tabular}{||c|rrr|rr|r||}
\hline\hline
&&&&&&\\
& \multicolumn{1}{|c}{Hart.}
&\multicolumn{1}{c}{HF1}
&\multicolumn{1}{c}{HF2}
&\multicolumn{1}{|c}{BHF}
&\multicolumn{1}{c}{DBHF}
&\multicolumn{1}{|c||}{Exp}\\
&&&&&&\\
\hline
&\multicolumn{6}{|c||}{}\\
&\multicolumn{6}{|c||}{Potential A}\\
&\multicolumn{6}{|c||}{}\\
&&&&&&\\
$\epsilon_{d5/2}$ [MeV] & -19.8 & -21.0 & -20.8 & -30.2 & -21.9 &
-14$\pm$2 \\
$\epsilon_{1s1/2}$ [MeV] & -15.4 & -13.7 & -14.1 & -24.5 & -13.8 &
-10$\pm$1 \\
$\epsilon_{d3/2}$ [MeV] & -14.5 & -13.2 & -14.2 & -16.5 & -10.2 &
-7$\pm$1 \\
&&&&&&\\
E/A [MeV] & -8.21 & -7.76 & -8.09 & -8.29 & -8.64 & -8.50 \\
$R_{ch}$ [fm] & 3.35 & 3.14 & 3.14 & 2.64 & 3.05 & 3.50 \\
&&&&&&\\
\hline
&\multicolumn{6}{|c||}{}\\
&\multicolumn{6}{|c||}{Potential C}\\
&\multicolumn{6}{|c||}{}\\
&&&&&&\\
$\epsilon_{d5/2}$ [MeV] & -15.3 & -16.9 & -16.7 & -21.0 & -16.5 &
-14$\pm$2 \\
$\epsilon_{1s1/2}$ [MeV] & -10.9 & -11.3 & -11.5 & -16.9 & -10.6 &
-10$\pm$1 \\
$\epsilon_{d3/2}$ [MeV] & -10.5 & -10.8 & -11.5 & -12.0 & -8.0 &
-7$\pm$1 \\
&&&&&&\\
E/A [MeV] & -5.83 & -5.80 & -6.14 & -5.06 & -5.91 & -8.55 \\
$R_{ch}$ [fm] & 3.44 & 3.31 & 3.32 & 2.87 & 3.21 & 3.50 \\
&&&&&&\\
\hline\hline
\end{tabular}
\end{center}
\end{table}
\clearpage
\begin{table}
\caption{Results of relativistic HF calculations on
\cafe\ using the OBE potential A. Further information see table 3.}
\label{tab:rescafe}
\begin{center}
\begin{tabular}{||c|rrr|r||}
\hline\hline
&&&&\\
& \multicolumn{1}{|c}{Hart.}
&\multicolumn{1}{c}{HF1}
&\multicolumn{1}{c}{HF2}
&\multicolumn{1}{|c||}{Exp}\\
&&&&\\
\hline
&&&&\\
$\epsilon_{d5/2}$ [MeV] & -24.6 & -29.0 & -27.2 & -20$\pm$1 \\
$\epsilon_{1s1/2}$ [MeV] & -18.7 & -19.5 & -20.2 & -15.8 \\
$\epsilon_{d3/2}$ [MeV] & -19.5 & -21.6 & -25.1 & -15.3 \\
&&&&\\
E/A [MeV] & -8.35 & -7.83 & -7.90 & -8.70 \\
$R_{ch}$ [fm] & 3.34 & 3.15 & 3.16 & 3.50 \\
&&&&\\
\hline\hline
\end{tabular}
\end{center}
\end{table}

\clearpage
%\section{Figures}
\begin{figure}[h]
%\special{isoscale d:/usr/fritz/coupac.hgl,\the\hsize 3.6in}
\vspace{14 cm}
\caption{Effective coupling constants for the exchange of a scalar
meson ($G_{\sigma}$) and vector meson ($G_{\omega}$) as a function of
the density. The coupling constants are derived from DBHF calculations of
nuclear matter employing the OBE potentials A (left part) and C (right
part). Various mean field approximations are considered: the Hartree
approximation, the Hartree-Fock approximation with only $\sigma$ and
$\omega$ exchange (HF1) and the complete model of section 2, also
including the pion (HF2)}
\label{fig:coupac}
\end{figure}
\begin{figure}[h]
%\special{isoscale d:/usr/fritz/nmhaa.hgl,\the\hsize 3.6in}
\vspace{14 cm}
\caption{Results for the binding energy per nucleon (upper part) and
the effective mass $M^*$ of the nucleon as a function of the Fermi
momentum in nuclear matter. Results obtained from DBHF calculations
(OBE potential A, solid line) are compared to those of Hartree calculations
in which the effective coupling constants are determined to reproduce
the DBHF results at $\rho =0.2 \rho_{0}$ ($k_{F}$=0.8, long dashed line)
and at $\rho =1.1 \rho_{0}$ ($k_{F}$=1.40,  dotted line).}
\label{fig:nmhaa}
\end{figure}

\begin{figure}[h]
%\special{isoscale d:/usr/fritz/nmhaa.hgl,\the\hsize 3.6in}
\vspace{14 cm}
\caption{The radial function for the large upper component $g(r)$ and
the small lower component $f(r)$ (both multiplied by $r$) for the
$0s_{1/2}$ Dirac spinor obtained in a relativistic HF calculation (HF1,
OBEPA) for \caf\ are compared to a harmonic oscillator spinor of
eq.(39) assuming no medium correction ($k_{F}=0$, dotted line) or as
predicted by nuclear matter at an average density (dashed line).}
\label{fig:wave}
\end{figure}

\end{document}